\documentclass[
    ,final            
  ]
  {aipproc}

\layoutstyle{6x9} 

\begin{document}

\title{The spin content of the proton}

\classification{14.20.Dh,12.39.Ba
}
\keywords      {Confined relativistic interacting quarks, cloudy bag model}

\author{F. Myhrer}{
  address={Dept. Physics \& Astronomy, Univ. South Carolina, 
Columbia, SC 29208}
}
%
%
%

\today

\begin{abstract}
Considerable progress has been made in our knowledge of the spin distribution within the proton. The recently measured limits on polarized gluons in the proton suggest polarized gluons contribute very modestly to the proton spin. We will show that a modern, relativistic and chirally symmetric description of the nucleon structure naturally explain the current proton spin data. 
Most of the ``missing" spin is carried by confined quark and antiquarks' angular momentum.  
\end{abstract}

\maketitle


\subsection*{Introduction}

During the 
past two decades there has been a 
concerted level of activity 
mapping the distribution of spin (and angular momentum) 
onto the quarks and gluons that compose the nucleon. 
This effort was sparked by the 
discovery by the 
European Muon Collaboration (EMC) 
of a proton ``spin crisis''~\cite{emc88}. 
EMC observed that the valence quarks carried only a very small 
fraction $\Sigma$ of the proton spin~\cite{emc88}. 
The published measurement of the fraction 
$\Sigma \simeq 14 \pm 9 \pm 21$\%, indicated that it could 
possibly be equal to zero whereas the 
Ellis-Jaffe sum rule, 
based on the non-relativistic quark model (NRQM), 
predicts that $\Sigma =1$. 
The unexpectedly small 
EMC value for $\Sigma$ generated a 
tremendous level of theoretical and experimental activity. 
Theoretically several well-known aspects of hadron structure 
were explored~\cite{mt88,hm88,st88} 
but none could generate such a small value for $\Sigma$. 
It was however quickly realized 
that the famous $U(1)$ axial anomaly 
could strongly influence the value of $\Sigma$ and that 
the proton might contain a large quantity of 
polarized glue, see e.g., Refs.~\cite{et88,la88,ar88,ccm88}
for a mathematically elegant formulation of this possible 
contribution to the proton spin. 
In addition, and in contrast to the NRQM treatment, 
since the u and d quarks in the proton behave  
relativistically, their angular momentums will contribute 
to the spin content of the proton. Schematically the proton 
spin content can be written as
\begin{eqnarray*}
\frac{1}{2} &=& \frac{1}{2} \Sigma + \Delta_{Glue} + L_z
\end{eqnarray*}
The very recent experimental measurements at 
CERN, DESY, JLAB, RHIC and SLAC  
%
%
have shown that $\Delta_{Glue}$ 
gives (at best) a small contribution to 
the proton spin~\cite{SPIN2008}. 
Furthermore, 
the accuracy of the measured $\Sigma$-value has increased  
and we now know that the sum of the quark helicities in the 
proton is about 1/3, 
\begin{eqnarray*}
\Sigma = 0.33 \pm 0.03 \pm 0.05 \; , 
\end{eqnarray*} 
which is considerably higher than the initial 
EMC suggestion. 
That polarized glue is not  
the explanation for the spin problem leads us 
to focus again on the suggestions  
which were based on physics that is 
more familiar to those modeling non-perturbative QCD. 
In particular, we~\cite{mt08} 
suggest that most of the missing spin of the 
proton must be carried as orbital angular 
momentum by the quarks and anti-quarks.   

Before explaining the key physics  which, 
when combined, appear to provide a natural 
explanation of $\Sigma$, 
in more detail, see e.g. Ref.~\cite{mt08},  
we present a quick 
summary  of these  phenomenologically 
well-established physics ``factors". They are:   
\begin{itemize}
\item The relativistic motion of the confined valence quarks 
\item The virtual excitation of anti-quarks 
in low-lying p-states generated by the 
one-gluon-exchange hyperfine interaction -- 
in nuclear physics terms this is called  
``an exchange current correction" 
\item The pion cloud of the nucleon's ``quark core".
\end{itemize}
These three pieces of physics, tested in 
many independent ways (see below), all 
have the effect of converting quark spin 
to orbital angular momentum. 
As explained in more detail below (and in a recent 
publication~\cite{mt08})  
the first 
reduces the spin by about one third, the 
second yields a reduction by an amount of 
order 0.15, and the third gives a multiplicative
reduction by a factor of order 0.7. 
Combining these physics effects  
reduces the fraction of the proton 
spin carried 
by its quarks to about one third, 
i.e. $\Sigma = (1-0.35 -0.15)\cdot 0.7 =0.35$.  
One important recent new observation, 
 based on a chiral analysis of data 
from lattice QCD~\cite{Young:2002cj},   
suggests very strongly that the 
pion cloud contributes very little to 
the  $\Delta$-N mass 
splitting.
This new finding 
provides the 
justification for combining the corrections to the 
spin sum arising from one-gluon-exchange to that from 
the pion cloud.  
We now present some details of these three major reduction factors,  
which lead to the small value of $\Sigma $.


\subsection*{Relativistic valence quark motion }

This effect on $\Sigma$ was  understood at the time of the EMC discovery. 
A spin-up, very light quark in an s-state, moving in a confined space, 
has a lower Dirac component in p-wave. 
The angular momentum coupling is such that for this component the spin 
is preferably down and reduces the ``spin content" of the valence quarks. 
In the bag model, for example, where the massless quark's ground state 
energy equals $\Omega /R \simeq 2.043/R$, the reduction factor 
$B= \Omega /3 (\Omega -1) \simeq 0.65$.  
The same factor reduces the value of $g_A$ from 5/3 to 
$\simeq$ 1.09 in a bag model 
and this value changes little if one uses 
typical light quark current masses. 
(In the discussion section we 
will briefly indicate how our model leads to a realistic 
$g_A$ value $\simeq 1.27$.) 
The quark energy, $\Omega /R$ is determined by the bag confinement condition 
that the quark current out of the spherical bag cavity of radius $R$ is zero, 
i.e. in Dirac's notation 
$\hat{r}\cdot \vec{j} = i\hat{r}\cdot \psi^\dagger \vec{\alpha} \psi = 0$ for 
$r=R$. Even in more modern relativistic models, 
where quark confinement is simulated 
by forbidding on-shell propagation through proper-time regularization, 
the reduction factor is very similar -- e.g., 
in Ref.~\cite{Cloet:2007em} $\Delta u + \Delta d$ is 0.67. 
The relativistic 
motion transfers roughly 35\% of the nucleon spin from quark spin 
to valence quark orbital angular momentum.

\subsection*{The one-gluon-exchange hyperfine interaction} 

%
%
\begin{figure}
 \includegraphics[height=.3\textheight]{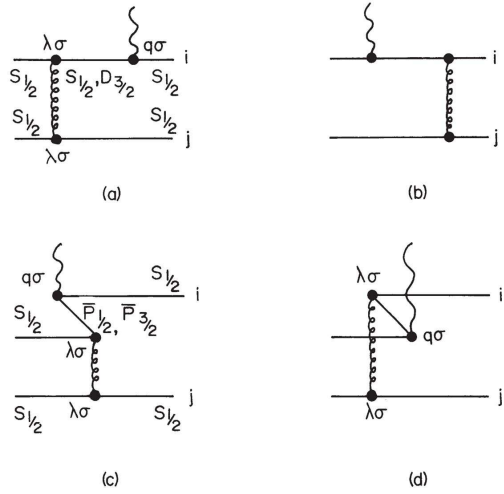}
 \caption{Illustration of the quark-quark hyperfine 
contributions which involve excited intermediate quark states. 
In the figures the external probe (top vertical wavy line) 
couples to the i'th quark which interacts with the second j'th quark via 
the effective confined gluon exchange. The intermediate quark propagator is 
evaluated as a sum over confined quark modes. Figs. (a) and (b) 
show the  three-quark intermediate states,  and (c) and (d) 
the  one anti-quark and four quarks intermediate states. 
The mode sum converges rapidly and the lowest 
anti-quark $P_{1/2}$ and $P_{3/2}$ modes dominate 
the mode-sum~\cite{Hogaasen:1988jd}.}
\end{figure}        

It is well established that the spin-spin interaction between quarks in a baryon, 
arising from 
the exchange of a single gluon, 
explains a major part of the mass 
difference between the octet and decuplet 
baryons -- e.g., the nucleon-$\Delta$ mass 
difference~\cite{De Rujula:1975ge}. 
This spin-spin interaction must therefore also play a role when an 
external probe interacts with 
the three-quark baryon state. 
In the context of spin sum rules, the probe couples to the all possible  
axial currents in the nucleon.
{\em That is, 
the probe not only senses a single quark current 
but a two-quark current as well}. 
The latter has an intermediate confined-quark propagator connecting the 
vertex of the probe and 
the spin-spin interaction between two quarks, and is similar to the exchange-current 
corrections which are 
well known in nuclear physics. 
In the  case of the two-quark current, investigated in detail in 
Ref.~\cite{Hogaasen:1988jd} using the MIT bag model, the confined quark 
propagator was written as a sum over quark eigenmodes and the dominant 
contributions were found to come from the 
intermediate p-wave anti-quark states. 
The primary focus of Ref.~\cite{Hogaasen:1988jd} was however the 
one-gluon-exchange corrections to the magnetic moments and semi-leptonic 
decays of the baryon octet (see below).

Myhrer and Thomas~\cite{mt88} realized the importance of this 
correction to the flavor singlet axial charge and hence to the proton 
spin, finding that it reduced the fraction of the spin of the nucleon 
carried by quarks,  calculated in the naive bag model by 0.15,  
i.e., $\Sigma \rightarrow \Sigma - 3G$~\cite{mt88}. 
The correction term,  
$G$, is proportional to $\alpha_s$ times certain 
bag model matrix elements~\cite{Hogaasen:1988jd}, where $\alpha_s$ 
is determined by the ``bare" nucleon-$\Delta$ mass difference.
Again, the spin lost by the quarks is compensated by orbital angular 
momentum of the  quarks and anti-quarks (predominantly $\bar{u}$ 
in the p-wave). 

\subsection*{The pion cloud }

The pion cloud is an effective description of the
 quark-antiquark excitations 
which are required by the spontaneous breaking of chiral symmetry in QCD. 
In fact, describing a physical nucleon as 
having a pion cloud which interacts with the valence quarks of the quark core 
(the  ``bare'' nucleon), in a manner dictated by the 
requirements of chiral symmetry, has been very successful 
in describing the properties of the 
nucleon~\cite{Theberge:1980ye,Thomas:1982kv,mbx:1981}. 
The cloudy bag model (CBM)~\cite{Theberge:1980ye,Thomas:1982kv} 
reflects this nucleon  description where  
the  nucleon consists of a bare nucleon, $| N>$, 
with a probability $Z \sim 1-P_{N\pi} -P_{\Delta \pi} \, \sim \, 0.7$, in addition to being described 
as a nucleon 
and a pion and a $\Delta$ and a pion, 
with probabilities $P_{N\pi} \sim 0.20-0.25 $ and 
$P_{\Delta \pi} \sim 0.05-0.10 $, respectively. 
The phenomenological constraints on these probabilities 
were discussed, e.g.~\cite{Speth:1996pz,Melnitchouk:1998rv}. 
%
%

The pion cloud effect on $\Sigma$ 
was investigated 
early 
by Schreiber and Thomas~\cite{st88}, 
who wrote the corrections to the spin sum-rules for the 
proton and neutron explicitly in terms of the probabilities 
above. 
To summarize 
Ref.~\cite{st88}: 
the pion cloud correction to the flavor singlet combination modifies the 
proton spin in the following manner:
\begin{equation}
\Sigma \rightarrow  \left(Z - \frac{1}{3} P_{N \pi} +\frac{5}{3} P_{\Delta \pi} \right) \Sigma \, .
\label{eq:pion}
\end{equation}
{}From the point of view of the spin problem, 
the critical feature of the pion cloud is that the 
coupling of the spin of the nucleon to the orbital angular momentum of 
the pion in the 
$N \pi$ Fock state favors a spin down nucleon and a pion 
with +1 unit of orbital angular momentum. 
This too has the effect of replacing quark spin by quark 
and anti-quark orbital angular momentum. 
Note that in the $\Delta \pi$ Fock component the spin of the 
baryon tends to point up (and the pion angular momentum down), 
thus enhancing the quark spin. 
Nevertheless, the wave function renormalization factor, $Z$, dominates, 
yielding a reduction by a factor between 0.7 and 0.8 for the range of 
probabilities quoted above.  

\subsection*{Other ``spin observables" affected by these corrections} 

Some of the quark hyperfine interaction (OGE) 
and the pion cloud corrections are illustrated in the following two Tables. 
A brief summary of these corrections are: 
\begin{description}
\item[The effective OGE]
(i) This correction is vital in the understanding of the measured strength of 
$\Sigma^- \to n + e^- + \bar{\nu}_e$, see Table 2. \\
(ii) Essential to explain the magnetic moment inequality 
$|\mu_\Lambda | < | \mu_{\Xi^-} |$, Table 1, \\
(iii) OGE introduces configuration mixing in baryon-octet ground 
states which can affect strongly the radiative decay of excited baryons 
to the ground state baryons. 
\item[Pion cloud]
(i) Crucial component of the neutron charge distribution. \\
(ii) Provides a large iso-vector contribution to the nucleon magnetic moment. \\ 
(iii) Gives the leading non-analytic chiral-loop corrections 
to nucleon observables. 
\end{description}

%

\begin{table}\caption{\label{one}\protect 
The baryon magnetic moments from the valence quark, the pion cloud
and the OGE  contributions as evaluated in~\cite{Hogaasen:1988jd}. 
}
$
\begin{array}{|l| |c| c|r|r|l|}
\hline 
{\rm Baryon \; \; } & {\rm Quark}& {\rm Pion} & {\rm OGE} 
& {\rm Mag  \; mom} &  {\rm PDG} - 2002 \\ 
\hline \hline 
\mu_p & \mu_q & \delta \mu_\pi &
0 & 2.79 & +2.79   \\
\hline 
\mu_n \; \;  &  - \frac{2}{3}\mu_q & 
- \delta \mu_\pi & \frac{2}{3}G^{\; \prime}   
& -1.92 & -1.91 \\ 
\hline 
\mu_{\Sigma^+} \; \;  & \frac{8}{9}\mu_q+\frac{1}{9}\mu_s &
 \frac{1}{2} \delta \mu_\pi^* & 0 
& 2.46 & +2.458 \pm 0.010 \\ 
\hline 
\mu_{\Sigma^-} \; \;  & -\frac{4}{9}\mu_q +\frac{1}{9}\mu_s & 
-\frac{1}{2} \delta \mu_\pi^* & -\frac{2}{3} G^{\; \prime }  
& -1.20 & -1.160 \pm 0.025 \\ 
\hline 
\mu_{\Xi^0} \; \;  & - \frac{2}{9}\mu_q - \frac{4}{9}\mu_s & 
\simeq 0  & \frac{2}{3} G^{\; \prime } 
& -1.20 & -1.250 \pm 0.014 \\ 
\hline 
\mu_{\Xi^-} \; \;  & \frac{1}{9}\mu_q - \frac{4}{9}\mu_s & 
\simeq 0  & -\frac{2}{3} G^{\; \prime }  
& -0.73 & - 0.6507 \pm 0.0025\\ 
\hline 
\mu_{\Lambda} \; \;  & - \frac{1}{3}\mu_s & 
0  & \frac{1}{3} G^{\; \prime }   
& -0.61 & - 0.61 \\ 
\hline 
\end{array} 
$ 
\end{table}

\begin{table}
\caption{\label{two}\protect 
The semi-leptonic decays of some baryons showing \underline{only} 
the valence quarks and OGE contributions~\cite{Hogaasen:1988jd}. 
Other corrections are implicit.  
}
$
\begin{array}{|l| |r| l|}
\hline 
{\rm B^\prime \to  B \; \; } & {\rm Quark + OGE } 
& 
{\rm PDG} - 2002 \\ 
\hline \hline 
n \to p  & \frac{5}{3}B + G \simeq 1.25 
& 1.27   \\
\hline 
\Sigma^- \to n  \; \;  & - \; \frac{1}{3} B - 2G \simeq -0.34  
& -0.34 
\\ 
\hline 
\Lambda \to p \; \;  & B \; \;\;\;\;\; \; \; \simeq 0.72 
& 
0.72 
\\ 
\hline 
\Xi^- \to \Lambda \; \;  & \frac{1}{3}B -G \simeq 0.19 
& 0.25 \pm 0.05 
\\ 
\hline 
\end{array} 
$ 
\end{table}


\subsection*{Epilogue }

When the pion cloud or gluon exchange corrections were first discussed, 
each one alone did not 
yield a correction large   
enough to resolve the ``spin crisis". 
{}Furthermore,  since the pion 
 might contribute a 
substantial fraction of the 
observed mass-splitting between the N and $\Delta$, 
to combine these two corrections 
would reduce the strength of OGE. 
However, in the last few years the  
chiral analysis of 
quenched and full lattice QCD calculations for the N and 
$\Delta$
masses as a function of quark 
mass~\cite{Young:2002cj}, 
concluded that pion effects likely contribute 
50 MeV or less of the 
300 MeV $N - \Delta$ mass difference. 
We can therefore without too large an error 
combine the one-gluon-exchange and pion cloud 
corrections in the quark spin sum. 
This combined correction will give a 
$\Sigma$ 
between 0.35 ($P_{N \pi} =0.25, P_{\Delta \pi}=0.05$) 
and 0.40 ($P_{N \pi} =0.20,P_{\Delta \pi}=0.10$) 
in excellent agreement with the modern data.
We note that 
$g_A$ is reproduced by the 
same corrections affecting the $\Sigma$. 
Relativity reduces 
the value of $g_A$ from 5/3 to 1.09 and OGE and  
the pion cloud as well as the center-of-mass 
corrections will increase the $g_A$ value from 
1.09 to 1.27. 
As seen in Table 1 
these corrections are cruzial in order to reproduce 
the baryon magnetic moments, 
i.e. the pion isovector cloud is an 
important correction 
to the nucleon magnetic moments and 
the OGE 
restores the ratio 
$\mu_p /\mu_n \simeq - 3/2$!\cite{Hogaasen:1988jd}. 

We have used a model of confined quarks 
to compute the matrix elements of the axial current to find a 
$\Sigma$ 
value 
relevant in the limit 
$Q^2 \to \infty$. 
Our result, $\Sigma \in (0.35,0.40)$, agrees 
very well with the experimental value. 
%
The flavor singlet spin 
operator however has a non-zero anomalous dimension, $\gamma$,  
%
%
and the observable $\Sigma$ should be  
renormalization group independent and  
gauge-invariant 
as defined by  Larin {\it et al.}~\cite{Larin:1994va}. 
%
Motivated by the observation that 
a valence dominated quark model can only match experiment for 
parton distribution functions at a 
low scale, e.g.~\cite{Gluck:1988ey},   
our value of the quark spin 
would need to 
be multiplied by a non-perturbative factor involving 
the QCD $\beta$--function and 
$\gamma$. 
%
This evolution factor 
is truly non-perturbative and its 
three-loop perturbation theory evaluation 
by Larin {\it et al.}~\cite{Larin:1994va}
is at best semi-quantitative.
Nevertheless, it is rigorously less than unity 
and at three-loops gave 
a value of order 0.6--0.8~\cite{Hogaasen:1995nn}. 
Multiplying the quark spin obtained above by this factor 
gives 
$\Sigma  \in (0.21,0.32)$, 
in excellent 
agreement with the current experimental value. 

In conclusion, the impressive experimental progress aimed at resolving 
the spin problem has established that the quarks 
carry about 1/3 of proton's spin and that the 
gluonic contribution appears to be  
too small to account for the difference. 
Instead, well known nucleon structure features like the pion cloud,  
the quarks' hyperfine interaction, and the 
relativistic motion of the confined quarks, appear  
to explain the modern value of $\Sigma$. 
These new insights make us conclude that the missing spin should be 
accounted for by the orbital angular momentum of the quarks and 
anti-quarks 
-- the latter associated with the 
pion cloud of the nucleon and the p-wave anti-quarks 
excited by the one-gluon-exchange hyperfine interaction. 
Exploring the angular momentum carried by quarks and anti-quarks 
is a major focus of the scientific program of the 12 GeV Upgrade at 
Jefferson Lab., and is 
a 
promising way 
to test the model ideas present here.
%


This work was supported in part by the NSF grant PHY-0758114.


\end{document}